\documentclass[runningheads,a4paper]{llncs}

\usepackage{amssymb}
\usepackage{graphicx}

\usepackage{url}
\urldef{\mailsa}\path|wouter@onzichtbaar.net|
\newcommand{\keywords}[1]{\par\addvspace\baselineskip
\noindent\keywordname\enspace\ignorespaces#1}

\begin{document}

\mainmatter  

\title{Introduction to a Temporal Graph Benchmark}

\titlerunning{Introduction to a Temporal Graph Benchmark}

%
%
\author{Wouter Ligtenberg
\and Yulong Pei}
\authorrunning{Introduction to a Temporal Graph Benchmark}

\institute{Department of Mathematics and Computer Science\\ 
 		TU Eindhoven, The Netherlands\\
\mailsa\\
}

%
%

\toctitle{Lecture Notes in Computer Science}
\tocauthor{Authors' Instructions}
\maketitle

\begin{abstract}
A temporal graph is a data structure, consisting of nodes and edges in which the edges are associated with time labels. To analyze the temporal graph, the first step is to find a proper graph dataset/benchmark. While many temporal graph datasets exist online, none could be found that used the interval labels in which each edge is associated with a starting and ending time. Therefore we create a temporal graph data based on Wikipedia reference graph for temporal analysis. This report aims to provide more details of this graph benchmark to those who are interested in using it.
\end{abstract}

\section{Introduction}
\label{intro}
A temporal graph is a data structure, consisting of nodes and edges in which the edges are associated with time labels. 
While many temporal graph datasets exist online, none could be found that used the interval labels in which each edge is associated with a starting and ending time. For this reason we generated several synthetic datasets and modified two existing datasets, i.e., Wikipedia reference graph~\cite{preusse2013structural} and the Facebook message graph, for analysis. In this report, we introduce the creation of the Wikipedia reference graph and analyze some properties of this graph from both static and temporal perspectives. This report aims to provide more details of this graph benchmark to those who are interested in using it and also serves as the introduction to the dataset used in the master thesis of Wouter Ligtenberg~\cite{ligtenberg2017tink}.

\section{Creation of the Temporal Graph Benchmark}
\label{creation}
This dataset was created from the Wikipedia refence set~\cite{preusse2013structural}. In this dataset edges represent Wikipedia articles and every edge represents a reference from one article to another. The edges have 2 values, one with the time stamp of the edge creation or the
edge removal, and one indicating if it was added or removed. We transformed this dataset into an interval set in which we connected the creation of an edge to the removal of an edge to create an edge with an interval. In this dataset we only included edges that were first added and later removed. If an edge is added but never removed we ignore it. Since this is the first real interval set we contacted both Konect\footnote{\url{http://konect.uni-koblenz.de/}} and Icon\footnote{\url{https://icon.colorado.edu}} to submit this dataset to their index for further research. Both instances accepted the dataset to their database. 

In details, this graph dataset contains the evolution of hyperlinks between articles of the Dutch version of Wikipedia. Each node is an article. Each edge is a hyperlink from one page to another (directed). In the dataset, each edge has a start and an end time, which allows us to look at the network in different points in time. The network under study has 67,8907 vertices and 472,9035 edges. The first edge is created on Tuesday August 28, 2001, and the last edge removal takes place on Sunday July 10, 2011. So, our data roughly has a time span of 10 years.

\section{Properties of the Temporal Graph Benchmark}
\label{properties}
In this section, we introduce the properties of this temporal graph benchmark in two different perspectives, i.e., static and temporal. For the static graph analysis, we ignore the time stamps and view this dataset as a static graph. Then we calculate three graph measures on this graph including degree, clustering coefficient and PageRank.

\subsection{Static Graph Analysis}

\subsubsection{Degree}
The average degree, average in-degree and average out-degree characterize the network in a way that it gives an indication of how many edges there are connected on average to a node. We also calculate the max degree as the comparison. The results are shown in Table~\ref{tb:degree}.

\begin{figure}
\begin{center}
\includegraphics[scale=0.3]{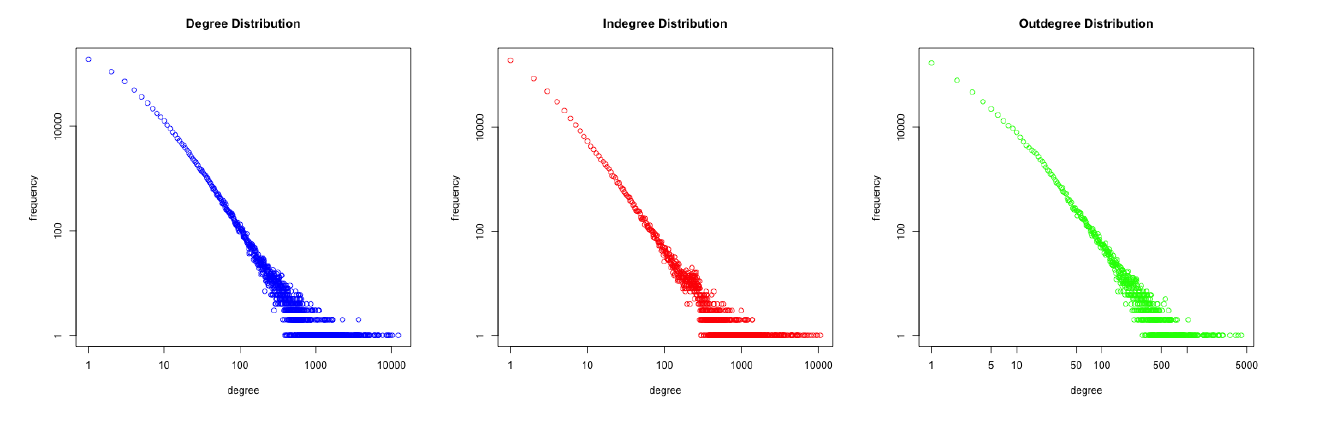}
\caption{Degree distributions of the static graph.}
\label{fig:degree} %
\end{center} 
\end{figure}

The degree distribution using a logarithmic scale for both the degree frequencies as the number of nodes including the degree distribution, in-degree distribution and out-degree distribution, are shown in Figure~\ref{fig:degree}. The curves indicate that these degree distributions follow the power law distribution.
\begin{table}[]
\centering
	\begin{tabular}{|c|c|c|c|}
	\hline
	& Degree & In-degree & Out-degree\\
	\hline
	average & 10.62 & 5.31 & 5.31\\
	\hline
	max & 12298 (n=394) & 10521 (n=394) & 4302 (n=306188) \\\hline
	\end{tabular}
\caption{Degree statistics of the static graph.}\label{tb:degree}
\end{table}

\subsubsection{Clustering coefficient}
How much nodes cluster together and form neighborhoods can be measured using the clustering coefficient. This measure quantifies the proportion of a node's neighbors to which the node is connected through triangles (three-node cycles). The clustering coefficient $C_i$ of node $i$ is defined as:
\begin{equation}
    C_i=\frac{2t_i}{k_i(k_i-1)},
\end{equation}
where $t_i$ is the number of triangles (three-node cycles) through $i$ and and $k_i$ is the degree of node $i$. $C_i = 0$ if none of the neighbors of a vertex are connected, and $C_i = 1$ if all of the neighbors are connected. Remark that the graph is considered as if it was undirected.

In this graph dataset, the average clustering coefficient is 0.114. For 28638 nodes $C_i = 1$, which means that for 4.22\% of the nodes all neighbors are connected. On the other hand, there are also 57.40\% of the nodes have $C_i = 0$. This means that these nodes are not part of any triangle at all.

\subsubsection{PageRank}
This measure ranks nodes by their importance. The PageRank value of a node is based on the nodes linking to and from that node. PageRank is mainly used by Google to rank websites by calculating the importance each website, but can also be used for other purposes such as rating Wikipedia articles.

\begin{figure}
\begin{center}
\includegraphics[scale=0.3]{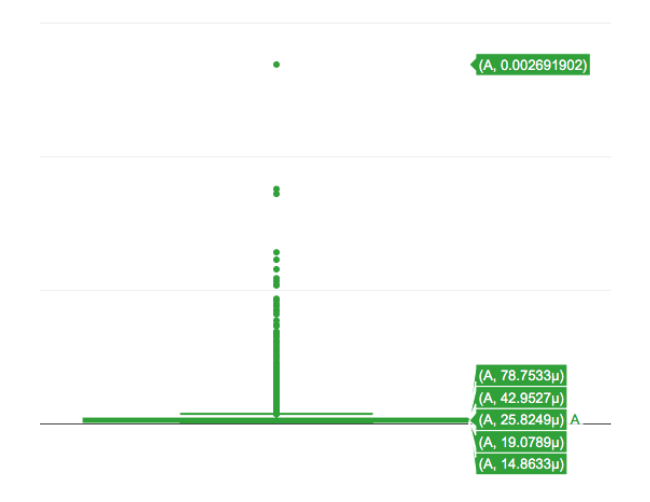}
\caption{PageRank of the static graph.}
\label{fig:pg} %
\end{center} 
\end{figure}

To visualize the results a boxplot was plotted, which is shown in Figure~\ref{fig:pg}. The figure clearly shows that only a few pages are classified as important, i.e. that have a high PageRank. While a lot of nodes/pages have relative small PageRank, i.e., less importance.

\subsection{Temporal Graph Analysis}
We use the snapshot-based method to analyze the temporal graph. Therefore the first step is choose the time window to divide the temporal graph into several snapshots. The time between the creation of the first edge, and the removal of the last edge covers a time period of 311,268,466 seconds, which is almost 10 years (9.87 years). For simplicity, we use the time window of 1 year in the temporal analysis to divide the temporal graph into 10 snapshots.

\subsubsection{Degree}
\begin{figure} 
\begin{center}
\includegraphics[scale=0.3]{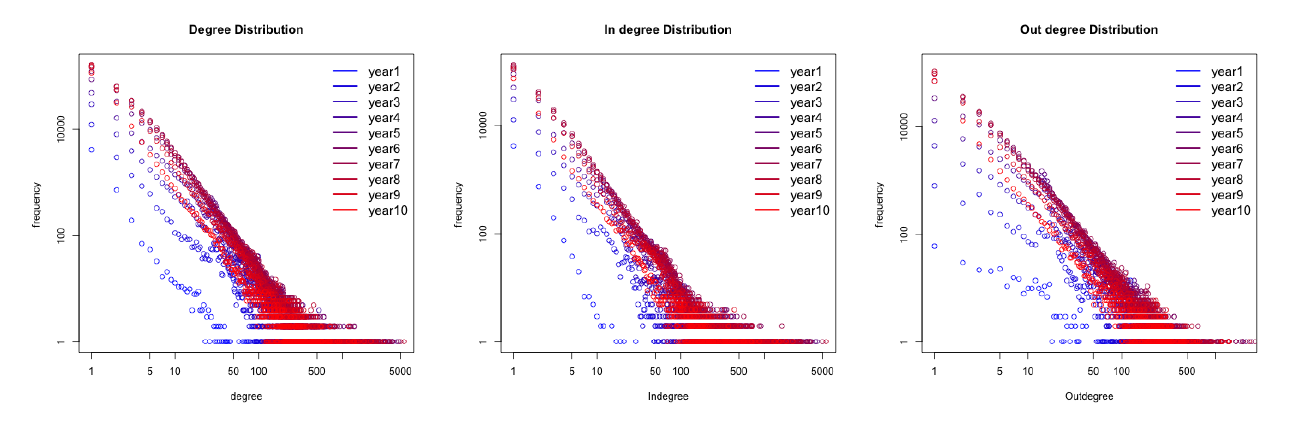}
\caption{Degree distributions of the temporal graph in different graph snapshots.}
\label{fig:tdegree} %
\end{center} 
\end{figure}

\begin{figure}
\begin{center}
\includegraphics[scale=0.27]{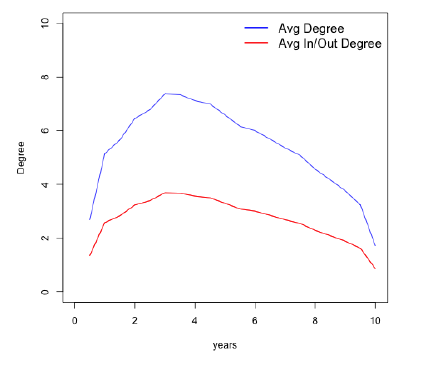}
\caption{Change of degrees in different graph snapshots.}
\label{fig:tcdegree} %
\end{center} 
\end{figure}

In Figure~\ref{fig:tdegree} the degree distributions per snapshot are visualized. Each year has its own color, ranging from blue (year 1) to red (year 10). We see a similar pattern for degree, in-degree and out-degree. In the first years, the graph shows a fairly large deviation from the power law. As the years pass, the graph develops to fit the power line better. The years 5 to 8 show a similar graph as we have seen in the static analysis of the graph. The last two years again start to deviate from this diagonal, although still keeping a fairly diagonal line. This behavior can be partly explained by the fact that the graph is growing at the beginning, and getting smaller again at the end. In all snapshots it seems to approximately hold that this network thus is a scale-free network. Moreover, the change of degrees in different graph snapshots is shown in Figure~\ref{fig:tcdegree}.

\subsubsection{Clustering coefficient}
\begin{figure}[h]
\begin{center}
\includegraphics[scale=0.27]{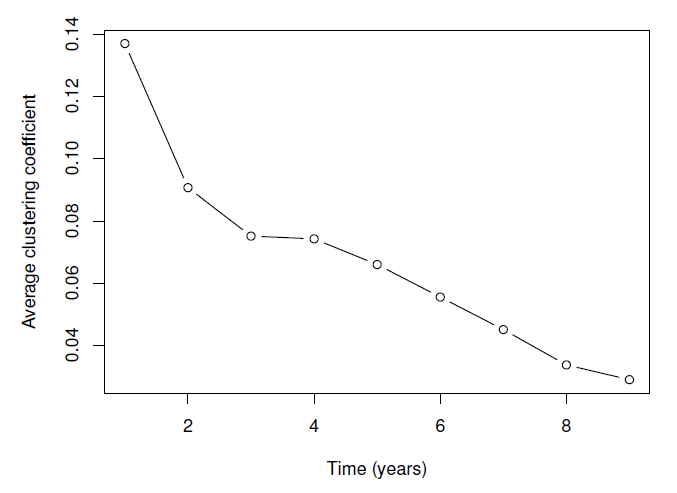}
\caption{Change of cluster coefficient in different graph snapshots.}
\label{fig:tcoeff} %
\end{center} 
\end{figure}

Figure~\ref{fig:tcoeff} shows the change of cluster coefficient of the temporal graph in different snapshots and the clustering does change over time. It is with relatively small steps, but we can conclude that the graph shows a downward trend. This trend might indicate that more Wikipedia pages tend to link to pages outside their general topic, which would lower the clustering coefficient. 

\subsubsection{PageRank}
\begin{figure}
\begin{center}
\includegraphics[scale=0.3]{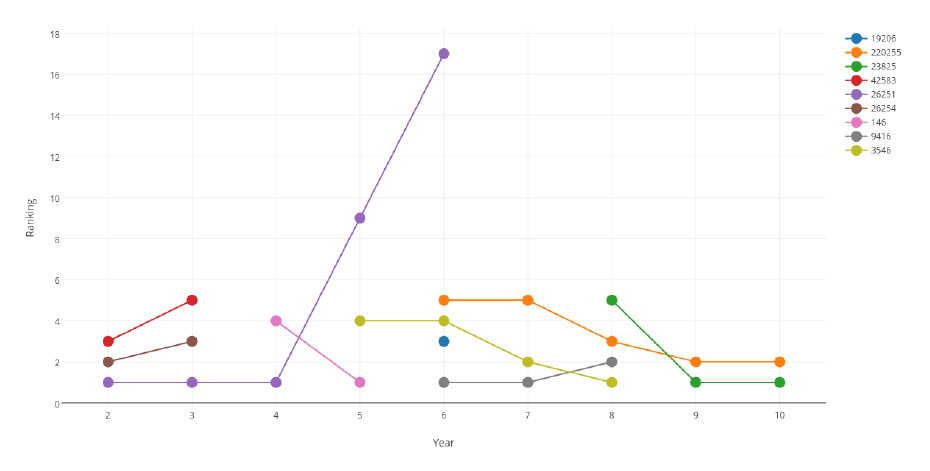}
\caption{Top nodes with highest PageRank in different graph snapshots.}
\label{fig:tpagerank} %
\end{center} 
\end{figure}

We use the PageRank to analyze the importance of nodes change over time. To visualize the results, we pick the top 30 nodes that scored the highest on PageRank and the results are shown in Figure~\ref{fig:tpagerank}. From the results, there is not a single most important node over the complete time period. Times change, inventions are done and new articles are written about new subjects. This is something
one would expect in a data file like this, and thus is confirmed by the data.

\section*{Acknowledgments}
Section~\ref{properties}, i.e., the analysis of graph properties, is from the advanced homework of 2\uppercase\expandafter{\romannumeral 2}D0 - Web Analytics course. We thank Evertjan Peer, Hilde Weerts, Jasper Adegeest and Gerson Foks (students of Group 12 in 2\uppercase\expandafter{\romannumeral 2}D0) for their efforts in the network analysis.

\end{document}